# Resistance Fluctuations in Composite Fermion State near Landau Level Filling Factor ν = 1/2 in A Square Quantum Well

*Jian Zhang[1,2], R. R. Du[*,1,3], J. A. Simmons[4], and J. L. Reno[4]*

[1]Department of Physics, University of Utah, Salt Lake City, Utah 84112
[2] Department of Physics, University of South Florida, Tampa, Florida 33620
[3] Department of Physics and Astronomy, Rice University, Houston, Texas 77251
[4]Sandia National Laboratories, Albuquerque, New Mexico 87185

We have studied the magnetotransport near Landau level filling factor ν = 1/2 in a GaAs-Al$_{0.3}$Ga$_{0.7}$As square quantum well (width 35 nm) in magnetic field up to 42 T and in a temperature range between 50 mK and 1.5 K. In low temperatures T < 200 mK, the R$_{xx}$ exhibits a deep, strongly temperature-dependent minimum centered at ν = 1/2. The concomitant R$_{xy}$ is not strictly linear with magnetic field (B) and the first derivative of R$_{xy}$ with respect to B shows a sharp cusp at ν = 1/2, indicating a non-classical Hall effect. We verify these characteristics are being single-layer origin by applying an in-plane magnetic field or electrostatic gates. The data may signal frustrations in the composite fermion state towards ν = 1/2 pairing in the quantum well where the electron layer thickness far exceeds the magnetic length.

High-mobility two-dimensional electron systems (2DES) in GaAs/AlGaAs heterostructures support a multitude of novel many-electron phases when it is subjected to an intense perpendicular magnetic field (B) and very low temperatures (T) [1,2]. Of particular interest are the quantum phases in a half-filled Landau level (Landau level index N) in a single-layer 2DES. Magnetotransport at half-fillings at different N in this system are found to exhibit completely different characteristics in terms of low temperature magnetotransport resistances, namely, the diagonal resistivity $\rho_{xx}$ and the Hall resistivity $\rho_{xy}$. Down to a typical temperature T ~ 20mK, such characteristics in the N = 0, 1, 2 have by and large conformed to the following pattern: 1) In the lowest LL, N = 0, the $\rho_{xx}$ at $\nu$ = 1/2, 3/2 show a weak temperature dependence, and the concomitant $\rho_{xy}$ is classical, *i. e.*, linear with magnetic field *B*. The transport features of 1/2 and 3/2 are well described by the formation of a composite fermion (CF) Fermi surface, which is compressible [3,4]. 2) In the first excited LL, N = 1, the $\rho_{xy}$ at $\nu$ = 5/2 (and $\nu$ = 7/2) shows quantized plateaus, and the $\rho_{xx}$ minimum shows activated transport associate with an energy gap [5-8]. The $\nu$ = 5/2, which is an even-denominator fractional quantum Hall effect (FQHE), and other states in N =1 LL, are probably the most interesting, and, enigma, due to their relevance to the non-abelian quasiparticles [9,10]. 3) In N = 2, magnetotransport at $\nu$ = 9/ 2, 11/2 are strongly anisotropic[11,12], a feature that can be attributed to a striped electronic phase or liquid crystal-like phases [13,14].

The physics underlying these rich phases and the phase transitions between them have been extensively studied in terms of pseudopotentials [15–17]. Rezayi and Haldane have shown by numerical calculations (exact diagonalization) that there exists a rather narrow window for stabilizing the quantized Hall states at half-fillings [17]. In particular, the pseudopotential ratio $V_1/V_3$ is found falling in this window for the formation of a paired FQHE state at 5/2 in GaAs/AlGaAs samples with realistic parameters. Increasing or decreasing this ratio would leads to a phases transition towards a compressible CF phase or an anisotropic phase. The latter scenario has been observed experimentally. For example, it is found that by applying an in-plane magnetic field, the 5/2 and 7/2 will transform to an anisotropic phase [18,19]. Overall, theories, as well as experiments in a few limited cases, have shown that the quantum phases and the phase transitions between them can be tuned by pseudo-potentials, which are in turn adjustable in real samples.

The question concerning the exact ground state at $\nu = 1/2$ in a single-layer QH sample, however, remains unsettled. Greiter, Wen, and Wilczek [20] proposed that the ground state at filling factors $\nu = 1/2$ is a BCS p-wave paired superconductor state, described by a Pfaffian wavefunction. Bonesteel [21] studied the stability of such paired states and found that the quantum fluctuation would eventually cause pair-breaking of the ground state, so the $\nu = 1/2$ would remain compressible (gapless) even at extremely low temperatures. Unlike the $\nu = 5/2$ state, which is believed to be described by Moore-Read wavefunction, the $\nu = 1/2$ state experiences strong fluctuations because the lowest LL is unscreened. It is suggested that the $V_1/V_3$ ratio can be influenced by the finite thickness of the 2DES, or, by a metallic gate close to the 2DES. Rezayi and Hadane have studied the influence of thickness on the phase diagrams for a number of cases [17]. More over, Park *et al.* proposed [22] that paired states could prevail at $\nu = 1/2$ in a thick 2DES. Specifically, they suggest that the regime of interest is being $\beta = \lambda / \ell_B > 5$, where $\lambda$ is the 2DES thickness and $\ell_B = \sqrt{\hbar/eB}$ is the magnetic length.

It is well known that, under appropriate conditions, the electron distribution profile in a wide QW can split into two separate layers due to Coulomb and exchange interactions. In this respect, it is worthwhile to review the phase diagram for the paired $v = 1/2$ FQHE observed in a bilayer context. In their systematic work on the wide QW experiments Suen *et al.* had identified a phase space range in which the $v = 1/2$ paired quantum Hall states were observed in magnetotransport [23]. The results are summarized in a phase diagram that is parameterized by $\alpha \equiv \Delta/(e^2/\varepsilon \ell_B)$ and $\beta = \lambda/\ell_B$, where $\Delta$ is the separation of the two lowest subbands in their QWs, and $\lambda$ here characterizes the spatial separation of two layers. Since the $\alpha$ represents the mixing of the symmetrical lowest subband wave function and the anti-symmetrical first-excited subband, and $\lambda/\ell_B$ the interlayer correlation, the two-component paired states are found to be stable only in a critical balance between the intra- and inter- layer correlations. In particular, such quantum liquids are most stable in an island in the phase space characterized by $\alpha \leq 0.1$ and $5 < \lambda/\ell_B < 7$. Outside this region other phases such as CF metallic states and insulating phases may prevail. Our experiments up to 42 T in a 35 nm QW shall span the parameters $5.5 < \lambda/\ell_B < 6.5$ whereas the $\alpha \sim 0.3$ representing a single–layer QH system. This region overlaps with the theoretically proposed regime for the Pfaffian state in a single-layer 2DES [22].

Our experimental investigation for $v = 1/2$ in a square quantum well (QW) is motivated by the above proposals. In order to reach a desired effective thickness $\beta$, we perform our experiments in ultrahigh magnetic field up to 42 T. The electron wavefunction in the well can be tuned by a pair of top-bottom gates. In addition, in-plane magnetic field can be applied by tilt-field experiments. Our specimens were fabricated from symmetrically Si-doped, 35 nm width GaAs-Al$_{0.3}$Ga$_{0.7}$As QW grown by molecular beam epitaxy on the (001) GaAs substrate, with a 70 nm spacer from either side. 2DES has a density $n_s \approx 3.5 \times 10^{11}/cm^2$ and a high mobility $\mu \geq 3 \times 10^6 cm^2/Vs$ at low temperatures. Total of 5 specimens from the same wafer were measured and similar data were obtained from all specimens. The specimens were patterned in either a 5 mm × 5 mm Van der Pauw square (with eight indium contacts diffused around the perimeter) or a 200 μm-width Hall bar. On gated Hall bar, front- and back- electrostatic gates were processed using a flip-chip technique [24]. Experiments were performed in NHMFL using a $^3$He-$^4$He dilution refrigerator combined with either a resistive or a hybrid magnet. A base temperature of 50 mK was attained in magnetic field up to 42 T. Low frequency (7 Hz) lock-in technique was employed and a mall current of 20 to 50 nA were used to avoid electron heating.

Our central finding for ultrahigh magnetic field transport near $\nu = 1/2$ in the QW is represented by the data shown in Fig. 1. Magnetotransport in 35 *nm* QW deviates drastically from the characteristic CF transport in thin electron layers. The $R_{xx}$ exhibits a sharp, strongly temperature-dependent minimum centered at $\nu = 1/2$. It is well documented that in heterojunctions, the $\nu = 1/2$ diagonal resistance is only weakly dependent on temperature [25], so our data in a wide QW is dramatically different as compared with that observed in the heterojunctions. More over, here the concomitant $R_{xy}$ is not strictly linear in B; its slope in the vicinity of $\nu = 1/2$ shows a sharp cusp.

We shall first focus on the T-dependence of diagonal resistance, emphasizing the following observations. 1) Temperature-dependent resistivity $\rho_{xx}$ exhibits different signs in $d\rho_{xx}/dT$, as is shown in the right-top panel. At a higher temperature range between ~

0.5 K < T < 1.5 K, the resistivity at $\nu = 1/2$, $\rho_{xx}(\nu = 1/2)$, decreases with T, whereas for T < 0.5 K, it is increasing with lowering T down to 50 mK, the lowers T in this experiment. In both regimes the $\rho_{xx}(\nu = 1/2)$ is approximately linear with log T. 2) On both sides of the half-filling, i.e., $\nu$ < 1/2 and $\nu$ >1/2, the $\rho_{xx}$ increases dramatically in lowering T, and the resistivity feature around $\nu$ = 1/2 can be characterized by a steep dip being developing in low temperatures. In the right-bottom panel the amplitude of resistivity difference, defined as $\Delta\rho_{xx} = \frac{1}{2}(\rho_1 + \rho_2 - 2\rho_{min})$, is plotted against T, where the data can be approximately fit into two lines, and two lines intersects at around 0.4 K. 3) The above features are so far observed in a relatively narrow filling factor range where the FQHE states are absent. Specifically, the strong T-dependent $\rho_{xx}$ were observed in the range 27 T < B < 32 T, bounded by clearly resolved 6/11 and 6/13 minima. For T < 200 mK, reproducible $\rho_{xx}$ fluctuations emerge in this magnetic field range, see, e.g., the small amplitude, peak-valley-like structure near 27.5 T. The pattern resembles FHQE minima but cannot be assigned to the standard CF series $\nu = p/(2p \pm 1)$ around $\nu = 1/2$,

This is not the first time that a steep $\rho_{xx}$ minimum is being observed at $\nu = 1/2$ in a high-mobility 2DES. Jiang et al [25] reported early on a remarkably sharp dip in resistivity at $\nu = 1/2$ and very strong positive magnetoresistance around it. However, in the single-interface GaAs-AlGaAs heterojunction studied in ref [25], the magnetoresistance around $\nu = 1/2$ is nearly temperature-independent, a fact that is consistent with the half-filled lowest Landau level being a CF metal. In this regard, our observation for $\rho_{xx}(\nu = 1/2)$ in a thick QW is distinct from the behavior of a CF metal. It is reasonable to think that the $\rho_{xx}$ strongly resemble early results for transport around the

ν=5/2 [5] in a heterojunction, or a developing $\nu = 1/2$ FQHE in a wide single QW [23]. In particular, the sign change of $d\rho_{xx}/dT$ at T ~ 0.5 K may indicate a developing minimum mixed with a raising back ground in this temperature range.

The Hall resistivity $\rho_{xy}$ at the $\nu = 1/2$ does not follow a strictly classical, *i. e.*, linear with *B*, pattern. We show in Fig. 2 the $\rho_{xy}$ measured in a Hall bar specimen, along with its derivative $B \cdot d\rho_{xy}/dB$, which is obtained numerically on the data of $\rho_{xy}$. The $\rho_{xy}$ ($B \cdot d\rho_{xy}/dB$) observed is perfectly anti-symmetrical (symmetrical) with respect to B = 0, indicating a negligible admixture between resistivity tensor components $\rho_{xx}$ and $\rho_{xy}$. From $B \cdot d\rho_{xy}/dB$, the FQHE series around $\nu = 1/2$, as well as that around ν = 3/2, are clearly observed. We reaffirm that the resistivity rule [26-29] is valid in the FQHE regime in our QW.

Our center of attention, however, is on the sharp cusp in $B \cdot d\rho_{xy}/dB$, observed at ~ $\pm 29.5T$, *i.e.*, ν = 1/2, as is shown in Fig. 2. Our data shows unequivocally that the Hall resistivity near ν = 1/2 is not classical, as opposed to the metallic behavior proposed for a homogeneous CF metal. More strikingly, the temperature-dependence of the data (inset) shows that such sharp cusp develops at a low temperature, below T ~ 200 mK. We notice also that the cusp is much narrower than the $\rho_{xx}$ minimum around $\nu = 1/2$, which indicates a strong deviation from the resistivity rule [26-29] in this filling factor range. Together with the strong temperature dependence of $\rho_{xx}$, we conclude that in low temperatures, the magnetotransport in our QW sample at $\nu = 1/2$ dramatically deviates from the metallic behavior commonly observed in a thin 2DES. The data may indicate frustrations in the CF state near $\nu = 1/2$ in a thick, single-layer, QW where the electron layer thickness far exceeds the magnetic length.

In the rest of the paper we shall focus on the evidences supporting the single-layer origin of the observed magnetotransport near $\nu = 1/2$. Fig. 3a shows simulated electron wavefunctions corresponding to respectively the subband $E_1$, $E_2$ at 4 K and in zero magnetic field using the materials parameters for the 35 nm QW. The calculations have taken into account the Coulomb and Hartee-Fock exchange energies. The subband splitting is $\Delta E = E_2 - E_1 \approx 82.5 K$, corresponding to $\alpha \sim 0.3$ hence a single-layer 2DES. The distribution of electrons along the z direction can be characterized by a finite thickness $\lambda = 29.5$ nm and we estimate $\beta \approx 6$ for $\nu = 1/2$ at 30 T. We also calculate pseudopotential ratio $V_1/V_3$ for our sample at $\nu = 1/2$. Comparing with the ideal Coulomb interaction, $V_1/V_3$ is about 10% smaller for our sample, whereas $V_1/V_3$ is about 17% smaller at $\nu = 5/2$ [17]. The repulsive part of the Coulomb interaction is reduced for our sample due to finite thickness, but the 2DES remain as a single-layer.

Fig. 3b and 3c show respectively the low B and the high B diagonal resistances in a gated Hall bar sample. By independently applying the gate potential to the front and the back gates, we were able to tune the symmetry of the electron profile along the z direction, while keeping the electron density constant. The beats in SdH shown in case (B) indicate the occupation of the second subband with a density of $3 \times 10^{10} / cm^2$ and the electron profile is symmetric; (A) and (C) shows the electrons occupy only the lowest subband as the QW being tuned to asymmetric. It is an important observation that the shape of the diagonal resistance does not depend on the QW symmetry. We further examine the transport near $\nu = 1/2$ in the gated QW by keeping the QW symmetric while tuning $n_s$, shown in Fig. 3d. In this plot the electron density varies from 2.73 to 3.91 $\times 10^{11}/cm^2$, corresponding to approximately $5.5 < \lambda/\ell_B < 6.5$. We observe that the resistance minimum becomes steeper with increasing $\lambda/\ell_B$.

In addition, we have checked the $\nu = 1/2$ features in a tilted magnetic field, and found that such features remain robust in all the tilt angles measured (up to tilt angle $36^o$, limited by total B = 42 T). On the contrary the known bilayer FQHE are extremely sensitive to either an asymmetry of the electron distribution, or a modest in-plane magnetic field [23]. We thus conclude that experimental observations exclude the transport features near $\nu = 1/2$ in our QW being a bilayer origin.

At this point there exist no satisfactory theoretical explanation for these dramatic observations. Interactions between CFs play important role and may influence magnetotransport near $\nu = 1/2$. For example, a logarithmic correction to Hall conductivity due to interaction of CFs is considered to account for modest positive magnetoresistance around $\nu = 1/2$ [30,31]. On the other hand, theory does not anticipate a strongly temperature-dependent, deep minimum in resistivity and a sharp cusp in concomitant Hall resistance around $\nu = 1/2$. We interpret the date as being indicative of competing CF phases in the vicinity of $\nu = 1/2$. Phenomenologically, the steep positive magnetoresistance around $\nu = 1/2$ resembles the magnetoresistance of a disordered modulated 2DES around B = 0 [32], where electrons aggregate and form clusters. It would be interesting to examine whether there exist emergent low energy phases, including novel inhomogeneous phases [7, 11-14], that are competing with the pairing phase in a thick QW.

In conclusion, in a QW 2DES with well width 35 nm, low temperature magnetotransport experiments in ultrahigh magnetic field up to 42 T have uncovered remarkable features near $\nu = 1/2$. At low temperature (< 200 mK), the $\rho_{xx}$ exhibits a sharp, strongly temperature-dependent minimum centered at $\nu = 1/2$. The concomitant $\rho_{xy}$ is not classical; its slope in the vicinity of $\nu = 1/2$ shows a sharp cusp. Moreover, the resistance around $\nu = 1/2$ shows reproducible peak-valley-like structures. These data deviate significantly from the characteristic composite fermions transport typical of thin 2DES, and may signal frustrations in the composite fermion state near $\nu = 1/2$ towards paring.

We acknowledge many helpful conversations with D. C. Tsui, H. L. Stormer, W. Pan, C. L. Yang, Y. W. Sue, J. K. Jain, Y. S. Wu, and E. H. Rezayi. JZ and RRD were supported by DOE Office of Sciences and NSF. The experiments were performed in NHMFL under the support of NSF and the State of Florida. We appreciate expert technical assistance by E. Palm and T. Murphy at NHMFL.

**References**

[1] *The Quantum Hall Effect*, edited by R. E. Prange and S. M. Girvin, 2nd ed., (Springer-Verlag, New York, 1990).
[2] *Perspectives in Quantum Hall Effects*, edited by S. Das Sarma and A. Pinczuk (Wiley and Sons, New York, 1998).
[3] J. K. Jain, Phys. Rev. Lett. **63**, 199 (1989).
[4] B. I. Halperin, P. A. Lee, and N. Read, Phys. Rev. B. **47**, 7312 (1993).
[5 ] R. Willett, J. P. Eisenstein, H. L. Stormer, D. C. Tsui, A. C. Gossard, and J. H. English, Phys. Rev. Lett. **59**, 1776 (1987).
[6] W. Pan *et al.*, Phys. Rev. Lett. **83**, 3530 (1999).
[7] J. P. Eisenstein et al, Phys. Rev. Lett. **88**, 076801 (2002).
[8] J. S. Xia et al, Phys. Rev. Lett. **93**, 176809 (2004).
[9] N. Read, Phys. Rev. Lett. **65**, 1502 (1990).
[10] G. Moore and N. Read, Nucl. Phys. B **360**, 362 (1991).
[11] M. Lilly et al, Phys. Rev. Lett. **82**, 394 (1999).
[12] R. R. Du et al, Solid State Commun. **109**, 389 (1999).
[13] M. M. Fogler, A. A. Koulakov, and B. I. Shklovskii, Phys. Rev. B **54**, 1853 (1996).
[14] E. Fradkin and S. A. Kivelson, Phys. Rev. B **59**, 8065 (1999).
[15] F. D. M. Haldane, Phys. Rev. Lett. **51**, 605 (1983).
[16] F. D. M. Haldane and E. H. Rezayi, Phys. Rev. Lett. **54**, 237 (1985).
[17] E. H. Rezayi and F. D. M. Haldane, Phys. Rev. Lett. **84**, 4685 (2000).
[18] W. Pan et al, Phys. Rev. Lett. **83**, 820 (1999).
[19] M. Lilly et al, Phys. Rev. Lett. **83**, 824 (1999).
[20] M. Greiter, X. G. Wen, and F. Wilczek, Phys. Rev. Lett. **66**, 3205 (1991).

[21] N. E. Bonesteel, Phys. Rev. Lett. **82**, 984 (1999).

[22] K. Park, Melik-Alaverdian, N. E. Bonesteel, and J. K. Jain, Phys. Rev. B **58**, R10167 (1998).

[23] Y. W. Suen *et al.*, Phys. Rev. Lett. **68**, 1379 (1992); **69**, 3551 (1992); **72**, 3405 (1994).
[24] M. V. Weckwerth, J. A. Simmons, *et al.*, Superlattices Microstruct. **20**, 561 (1996).
[25] H. W. Jiang *et al.*, Phys. Rev. B **40**, 12013 (1989).
[26] A. M. Chang and D. C. Tsui, Solid State Commun. **56**, 153 (1983).
[27] T. Sajoto *et al.*, Phys. Rev. B **41**, 8449 (1990).
[28] W. Pan et al, Phys. Rev. Lett. 95, 066808 (2005).
[29] S. H. Simon and B. I. Halperin, Phys. Rev. Lett. **73**, 3278 (1994).
[30] L. P. Rokhinson and V. J. Goldman, Phys. Rev. B **56**, R1672 (1997).
[31] D. V. Khveshchenko, Phys. Rev. B **55**, 13817 (1997).
[32] see e.g., S. Hugger, T. Heinze, and T. Thurn-Albrecht, arXiv: 0806.3896.

**Figure Captions**

Fig.1 The diagonal resistance, $R_{xx}$, and Hall resistance, $R_{xy}$, near Landau level filling factor $\nu = 1/2$ measured in a 35 nm GaAs/AlGaAs QW sample at various temperatures in magnetic field up to 33 T. Inset: a) the temperature dependence of the resistance at $\nu = 1/2$; b) the temperature dependence of the strength of the valley-peak around $\nu = 1/2$.

Fig. 2 First derivative of Hall resistance with respect to magnetic field shows a sharp minimum at $\nu = 1/2$. The inset shows temperature dependence of the derivative; notice that a sharp cusp develops below T ~ 200 mK.

Fig. 3 (a) Simulations of the electron wavefunction, electron density distribution, and energy separation, $\Delta_{12} = E_2 - E_1$, of the first excited subband from the ground state band for the 35 nm QW. (b) Using a front gate and a back gate the electron distribution in the QW is tuned towards front (A); back (C); and symmetrical (B); traces are shifted vertically for clarity. (c) The $\nu =1/2$ magnetoresistance minimum is not dependent on the electron distribution, indicating a single-layer behavior. (d) Using a front gate and a back gate the electron distribution is tuned to symmetrical at various densities; the relative strength of the $\nu =1/2$ minimum is shown to be enhanced as the electron density increases. The electron densities represented are respectively 2.73, 3.10, 3.47, 3.78, and 3.91 $\times 10^{11} / cm^2$. For comparison, $R_{xx}$ is normalized by the value at $\nu =1/2$ and the magnetic field axis is scaled with an electron density of $3.47 \times 10^{11} cm^2$ ($V_g = 0$).

**FIG. 1**

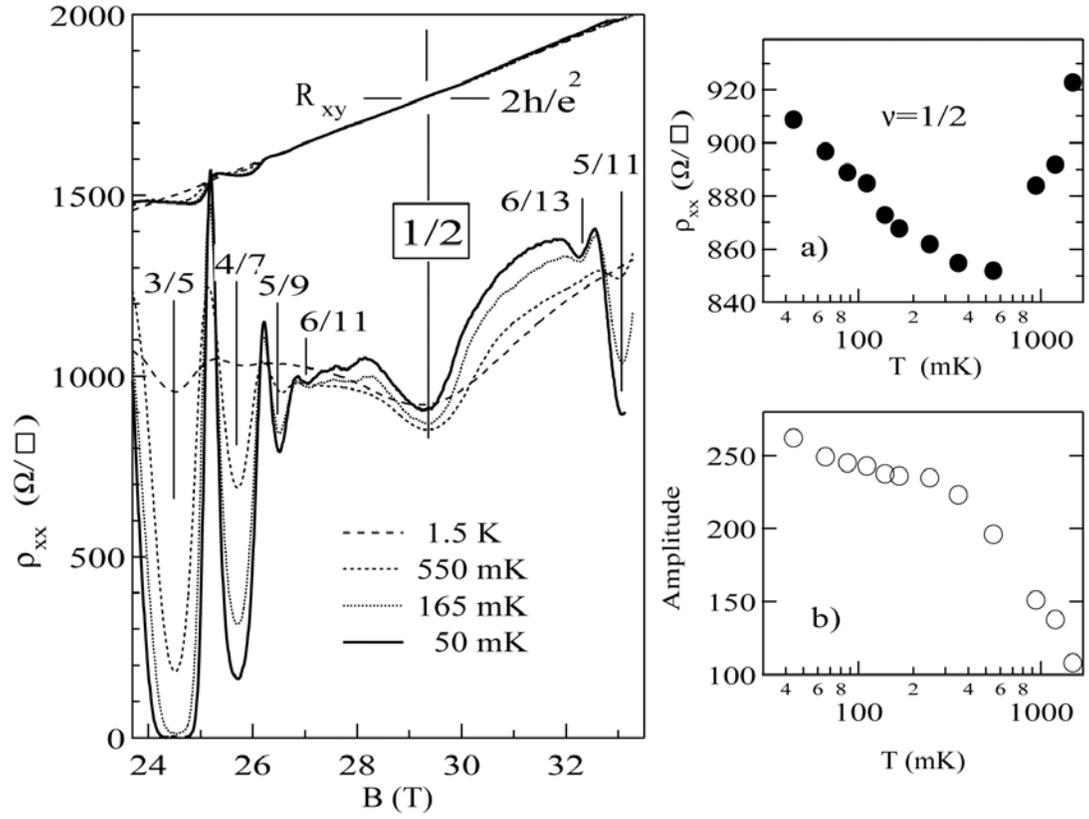

**FIG. 2**

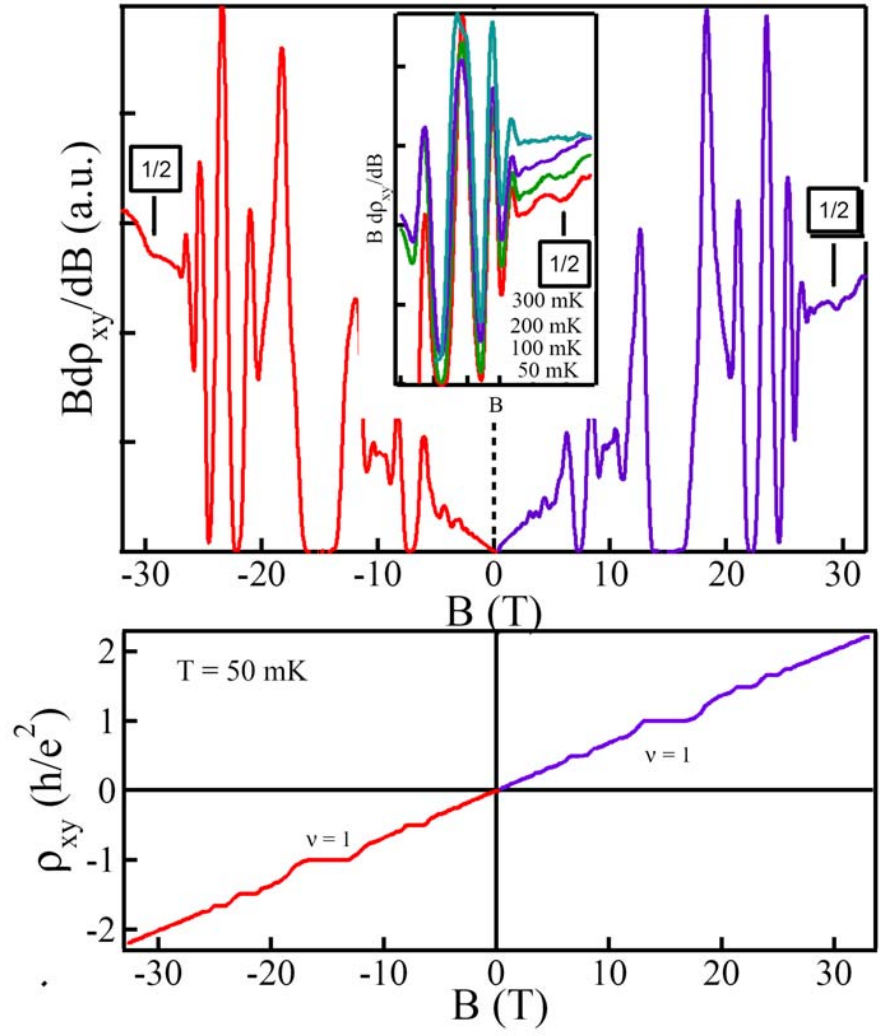

**FIG. 3**

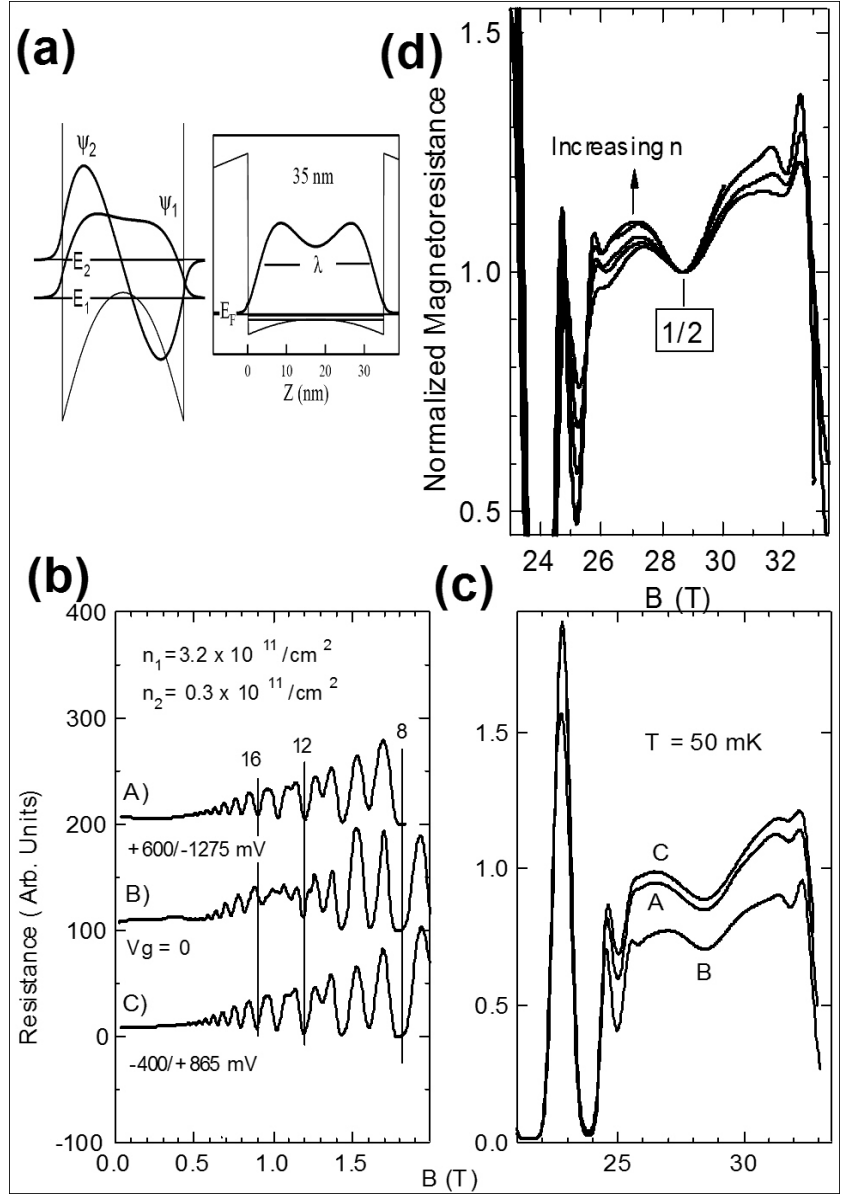